\documentclass[letterpaper, 10 pt, conference]{ieeeconf}  

\IEEEoverridecommandlockouts                              

\usepackage{graphicx}
\let\labelindent\relax
\usepackage{enumitem}
\title{\LARGE \bf
Learning Complex Multi-Agent Policies in Presence of an Adversary
}

\author{Siddharth Ghiya$^{1}$ and Katia Sycara$^{2}$
\thanks{*This work was supported by DARPA OFFSET award HR00111820029}
\thanks{$^{1}$Siddharth Ghiya is with Robotics Institute, School of Computer Science,
        Carnegie Mellon University, Pittsburgh, PA 15232, USA
        {\tt\small sghiya2@cs.cmu.edu}}%
\thanks{$^{2}$Katia Sycara is with Robotics Institute, School of Computer Science,
        Carnegie Mellon University, Pittsburgh, PA 15232, USA
        {\tt\small sycara@cs.cmu.edu}}%
}

\begin{document}
\maketitle
\thispagestyle{empty}
\pagestyle{empty}

\begin{abstract}
In recent years, there has been some outstanding work on applying deep reinforcement learning to multi-agent settings. Often in such multi-agent scenarios, adversaries can be present. We address the requirements of such a setting by implementing a graph-based multi-agent deep reinforcement learning algorithm. In this work, we consider the scenario of multi-agent deception in which multiple agents need to learn to cooperate and communicate to deceive an adversary. We have employed a two-stage learning process to get the cooperating agents to learn such deceptive behaviors. Our experiments show that our approach allows us to employ curriculum learning to increase the number of cooperating agents in the environment and enables a team of agents to learn complex behaviors to successfully deceive an adversary. \\

\end{abstract}

\section{INTRODUCTION}
Real world environments often consist of multiple agents which need to either collaborate or compete with each other to perform their task successfully. For example, an environment of self driving cars can be a multi-agent environment in which multiple autonomous cars need to collaborate and communicate with each other for effective decision making. multi-agent reinforcement learning (MARL) can be used to train a team of agents in such scenarios by maximising a reward function. However, most of the existing frameworks in this domain focus only on training collaborative policies for multiple homogeneous agents. Our work is specifically focused on the task of learning transferable collaborative policies in the presence of an adversary in the environment. For our experiments, we have considered the standard multi-agent task of deception in which multiple agents need to collaborate with each other in order to deceive an adversary. In such a scenario, designing heuristic guided behaviors for a team of agents is not a trivial task. We have adopted the framework of centralised training and decentralised execution in which only a single neural network is trained using PPO\cite{schulman2017proximal}. During testing however, every agent acts independently in the environment and multiple agents need to communicate with each other in order to reach consensus.

Through this work, we have made the following contributions:
\begin{itemize}
    \item We have modified a graph neural network based multi-agent reinforcement framework proposed by \cite{agarwal2019learning} to incorporate an opponent observation module. The proposed framework enables a group of collaborative agents to come up with behaviours to successfully deceive an adversary present in the environment.
    \item We have proposed a unique two stage training procedure using curriculum learning to enable a group of agents to learn complex cooperative policies in the presence of an adversary in the environment.
\end{itemize}

\section{RELATED WORK}
There has been a recent surge in the application of deep learning to reinforcement learning. Researchers have come up with various off-policy\cite{mnih2013playing}\cite{lillicrap2015continuous} and on-policy\cite{schulman2015trust}\cite{schulman2017proximal}\cite{mnih2016asynchronous} algorithms and have demonstrated super human performance. Multi-agent reinforcement learning is one of the more widely studied topics in the field of reinforcement learning. Independent Q-Learning\cite{DBLP:journals/corr/TampuuMKKKAAV15} represents some of the earliest works in this field. In Independent Q-Learning, agents in the environment are trained with the assumption that the other agent is part of the environment. Naturally, this fails if we increase the number of agents in the environment due to non-stationarity in the environment.

More recently, there has been some work in MARL which can be classified under the centralised training and decentralised execution paradigm. \cite{DBLP:journals/corr/LoweWTHAM17} proposed to modify the critic to evaluate the value of the next state conditioned on the actions of all the agents in the environment. They reasoned that training such a critic helped to counter the non stationarity introduced due to multiple actors in the environment. \cite{DBLP:journals/corr/FoersterFANW17} pointed out the problem of credit assignment in multi-agent reinforcement learning problems and proposed a method to assign credit to the action taken by an agent in such a setting. \cite{DBLP:journals/corr/SunehagLGCZJLSL17} proposed Value Decomposition Networks (VDN) to decompose the team value function to agent specific value functions. \cite{DBLP:journals/corr/abs-1803-11485} further built on the idea of VDN and proposed QMIX with an additional mixing network. The weights of such a mixing network are produced by another set of hyper-networks. They argue that their method represents a richer class of action-values functions. In most of these works, a centralised critic is maintained during the training process which accounts for the non-stationarity in the environment and hence they fall under the paradigm of centralised training with decentralised execution. In all of these works, no communication is assumed between the agents.

Researchers have also proposed communication between agents in multi-agent systems in order to encourage cooperation. \cite{foerster2016learning} proposed Differentiable Inter Agent Learning(DIAL) and Reinforced Inter Agent Learning(RIAL) as communication modules to enable agents to communicate with each other to be able to successfully collaborate with each other. \cite{mordatch2017emergence} showed the emergence of grounded language communication between agents. \cite{sukhbaatar2016learning} proposed CommNet which used continuous communication for cooperative tasks.

Some of the existing methods in research don't explicitly make assumptions about the type of other agents in the environment. In this context, there has been work where an agent actively tries to modify the learning behaviour of the opposite agent \cite{foerster2017learning} or employs recursive reasoning to reason about the behaviour of other agents in the environment \cite{wen2018probabilistic}. \cite{alshedivat2017continuous} used meta-learned policies to enable agents to adapt to different kinds of other agents in the environment. They were able to demonstrate such adaptive behaviours of the agents in both cooperative and competitive settings. \cite{grover2018learning} proposed learning policy representations of other agents in the environment in order to make more informed decisions. \cite{he2016opponent} developed Deep Reinforcement Opponent Networks (DRON) to model the opposite agents in the environment. \cite{Albrecht_2018} have done a detailed study about the available literature in the field of opponent modelling in multi-agent systems.

\cite{das2018tarmac} \cite{agarwal2019learning} proposed communication between the agents using dot-product attention mechanism. In this work, we have built on the framework proposed by \cite{agarwal2019learning} and modified it so that a team of agents can collaborate and come up with strategies to deceive an adversary present in the environment.

\begin{figure}[t]
\begin{center}
\setlength{\unitlength}{0.012500in}%
\thicklines
\includegraphics[scale=0.4]{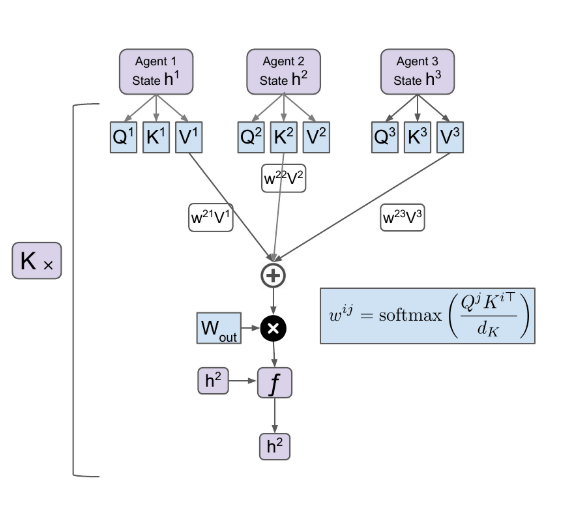}
\end{center}
\caption{Each agent receives $V^{j}$ and $Q^{j}$ from other agents in the environment. It then uses it's own key $K^{i}$ to produce attention it needs to pay to the message it received from agent $j$. It then aggregates over the messages it received from the other agents in the environment. This aggregated embedding is used to produce the action the agent takes in the current time step. This is similar to the architecture used by \cite{agarwal2019learning}.}
\label{message_passing} 
\end{figure}

\section{METHOD}
In \cite{agarwal2019learning}, authors have proposed modelling the environment as a graph of entities. In such a graph, nodes would represent an entity and and the interaction between two entities can be represented as edges. An example of an environment being modelled as a graph can be an environment of a self driving vehicle. In such a graph, cars, buildings and pedestrians would be modelled as nodes in the graph and their interactions such as communication between self driving vehicles would be modelled as edges in the graph.

Following a similar approach, in this work, we have also modelled the agents and landmarks in the environment as nodes in a graph, $G=(V,E)$, where V represents vertices and E represents edges. The agents learn to communicate important information about their surroundings to other agents. Since we have two types of teams in the environment, only agents in the same team are allowed to communicate with each other. All of the agents can observe the static entities in the environment(landmarks). Every agent has four modules : Agent State Encoder, Environment Encoder, Opponent Encoder and Inter Agent Communication Module. All of these modules have been explained in detail below:

\subsection{Agent State Encoder}
Each agent $i \in V$ can observe it's own state which includes it's own position and velocity. The agent forms its own state encoding $U^{i} = f_{a}(X^{i})$ using a learnable differentiable encoder, $f_{a}$. 

\subsection{Environment State Encoder}
Each agent is also able to observe all the entities in the environment and uses a Graph Neural Network to produce a fixed size embedding $E^{i}$. Note that this embedding is independent of the number of entities present in the environment. First, an agent uses an entity encoder function $f_{e}$ to produce an entity embedding, $e^{l}_{i}=f_{e}(X^{l}_{i})$ where $X^{l}_{i}$ is the position of entity $l \in V$ with respect to agent $i$. Then the agent produces a fixed size embedding $E^{i}$ by applying dot product attention mechanism over entity embeddings. $E^{i}$ can be intuitively understood as a representation of the environment of an agent. An important characteristic of using such technique is that the produced final embedding $E^{i}$ is invariant to the number of entities present in the environment. It is also invariant to the order in which the agent observes the entities.

\begin{figure}[t]
\begin{center}
\setlength{\unitlength}{0.012500in}%
\thicklines
\includegraphics[scale=0.25]{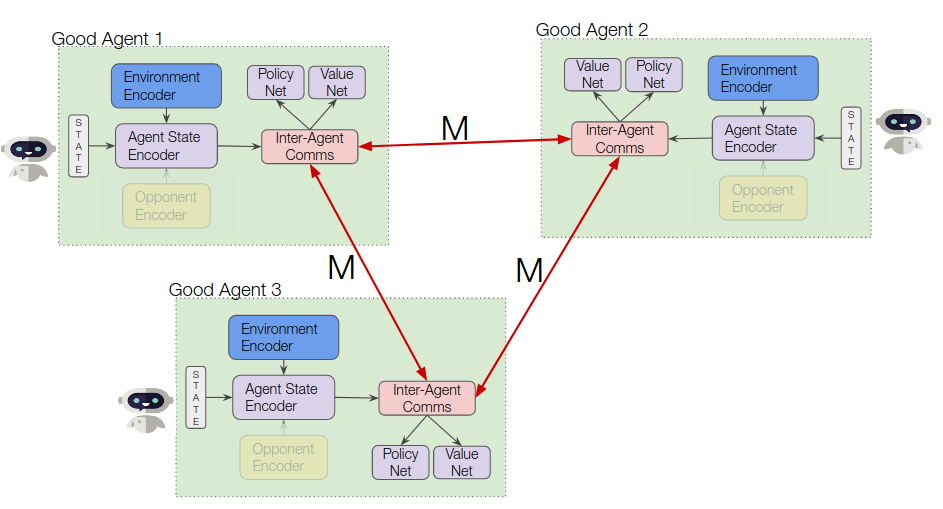}
\end{center}
\caption{Inter agent communication when we have scenarios with only homogeneous agents learning to collaborate with each other. The opponent encoder module has been disabled here.}
\label{homogeneous_architecture} 
\end{figure}

\subsection{Opponent State Encoder}
There can be scenarios where we have different teams of agents in the environment. In such scenarios, we also have an opponent encoder module which similarly takes in the state information of the opponent agents and produces an embedding reflecting the opponent information. Every agent uses an opponent encoder function $f_{o}$ to produce an opponent embedding $e^{o}_{i}=f_{o}(X^{o}_{i})$ where $X^{o}_{i}$ is the position of opponent $o \in V$ with respect to agent $i$. After producing an opponent encoding for all the opponents in the environment, a fixed size embedding $O^{i}$ is computed by applying dot product attention mechanisms over opponent embeddings. Similar to the environment embedding $E^{i}$, $O^{i}$ is invariant to the number of opponents in the environment and to the order in which the agent observes them.

\begin{figure}[t]
\begin{center}
\setlength{\unitlength}{0.012500in}%
\thicklines
\includegraphics[scale=0.241]{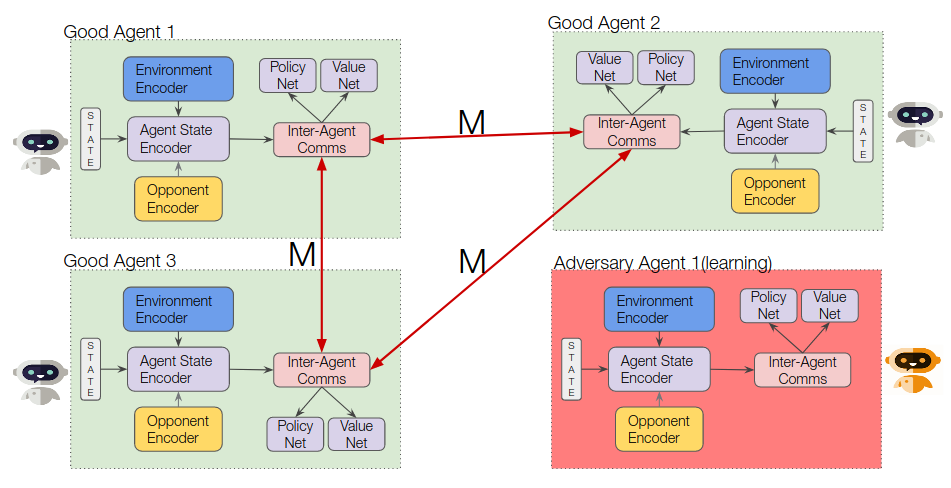}
\end{center}
\caption{Inter agent communication when we have scenarios with different teams of agents. We have an opponent encoder module}
\label{heterogenous_architecture_learning} 
\end{figure}

We can have scenarios where we only have homogeneous agents operating in the environment or we can have different teams of agents in the environment. In case of homogeneous agents operating in the environment, every agent is allowed to communicate with every other agent in the environment (Figure \ref{homogeneous_architecture}). On the contrary, if we have different teams of agents in the environment then the agents are only allowed to communicate with other agents in the same team (Figure \ref{heterogenous_architecture_learning}).

\subsection{Inter Agent Communication Module}
After computing its state encoding $U^{i}$, its environment encoding $E^{i}$, and an its opponent encoding $O^{i}$, an agent uses them to produce a message. It also uses them to choose how much attention it wants to pay to the messages it receives from other agents. First the agent concatenates $U^{i}$, $E^{i}$ and $O^{i}$ to produce $h^{i}$. Here, $h^{i}$ represents an agent's understanding of its own environment.

Using $h^{i}$, Each agent produces a key $K^{i} = W_{K}h^{i}$, a value $V^{i} = W_{V}h^{i}$ and a query $Q^{i} = W_{Q}h^{i}$. It then proceeds to send the computed $V^{i}$ and $Q^{i}$ to all the other agents in the environment. It also receives $V^{j}$ and $Q^{j}$, $j \in V-\{i\}$, where $j$ is every other agent in the environment. It then uses dot product attention method to calculate the attention it needs to pay to the message of agent $j$ at every time step. This method of calculating attention and aggregating information received from other agents is described in more detail in Figure \ref{message_passing}. After message passing, every agent updates its embedding $h^{i}  $. Now each agent passes it's hidden embedding, $h^{i}$ through another neural network which produces a distribution over the actions that the agent can take. At each time step, all the agents calculate their own action in a decentralised manner and are given a single reward for their collective set of actions. This reward is then used to train the agents using PPO \cite{schulman2017proximal}. One important implementation detail is that we assume that the agents in the same team share all the learnable parameters of agent state encoder network, environment state encoder network, opponent state encoder network, inter agent communication module and final policy network. This means that our work falls under the paradigm of centralised training and decentralised execution.

\section{EXPERIMENTS}

\begin{figure}[t]
\begin{center}
\setlength{\unitlength}{0.012500in}%
\thicklines
\includegraphics[scale=0.25]{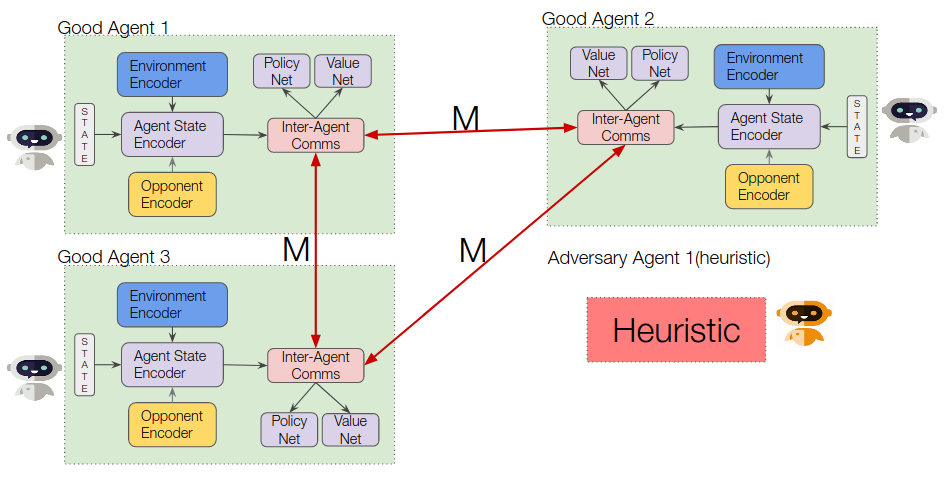}
\end{center}
\caption{Inter agent communication when we have scenarios with an heuristic adversary.}
\label{heterogenous_architecture_heuristic} 
\end{figure}

In this work, we have worked on the standard swarm robotic task of deception. The deception environment has been implemented in the Multi-Agent Particle Environment \cite{DBLP:journals/corr/LoweWTHAM17} where the agents can move around in 2D space following a double integrator dynamics model. Every agent produces an action in the form of acceleration along the X or Y direction.

\subsection{Environment Description : Deception}
In the task of deception, we can have a team of N agents trying to protect high value target among N targets. The agents needs to learn to spread out to cover targets in order to confuse an observing adversary as to which of the target is the most important. There are two teams of agents in the environment:
\begin{itemize}[leftmargin=*]
    \item \textbf{Good Agents}: They can observe all the landmarks in the environment and know which landmark is the target landmark. In our work, the good agents are collectively learning to collaborate with each other and to come up with strategies to deceive the observing adversary.
    \item \textbf{Adversary Agent}: In this work, we have only presented results with one adversary agent in the environment. We have tried experiments with both learning and heuristic adversary. Basically, an adversary needs to infer the target location from the motion of the good agents. If learning, the adversary can see all the landmarks but does not know which landmark is the target landmark. If heuristic, the adversary moves towards landmark with closest good agent. In this work, results have been presented with only a heuristic adversary.
\end{itemize}

Apart form the agents in the environment, we also have landmarks as entities in the environment. The kind of landmarks in the environment are as follows:
\begin{itemize}[leftmargin=*]
    \item \textbf{Target Landmark}: Target landmarks are the high value landmarks in the environment. The good agents in the deception environment can observe all the landmarks and know which landmark is the target landmark. Adversary agents do not know which landmarks are the target landmarks. The adversary agent needs to infer the target landmark by observing the good agents while the good agents need to come up with strategies to deceive the adversary.
    \item \textbf{Non Target Landmark} : Non target landmarks are the other landmarks in the environment. Both good agents and the adversary agents can observe all the landmarks in the environment.
\end{itemize}

\begin{figure}[t]
\begin{center}
\setlength{\unitlength}{0.012500in}%
\thicklines
\includegraphics[scale=0.5]{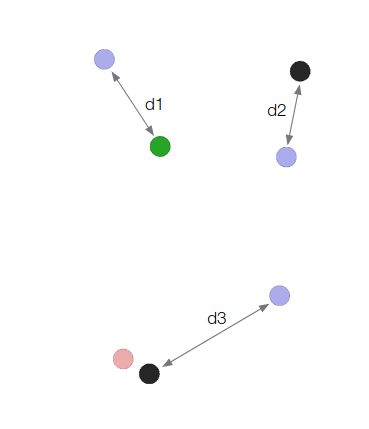}
\end{center}
\caption{This figure provides a visual representation of the deception environment. In this image: \textbf{Green colored dot} represents the target landmark, \textbf{Black colored dots} represent the other landmarks, \textbf{Blue colored dots} represent the good agents and \textbf{Red colored dot} represents the adversary agent. If we have a heuristic adversary, it tries to move towards the landmark with closest good agent or the landmark corresponding to $min(d1,d2,d3).$ Notice that the good agents know which landmark is green while the adversary agent needs to infer this information by observing the good agents.}
\label{deception}
\end{figure}

\subsection{Reward Description}
In this section, we describe the reward formulation which we have used for our experiments. Reward formulation in reinforcement learning should be done according to the behaviour which we want the learned agent to perform. In our case, we will describe different reward configurations and how we used them to obtain desired behaviour. The reward which we have used are based on distance and are continuous. Since the reward configuration is continuous and given at every time step, learning different behaviours becomes easier for the agents. We employed curriculum learning to make agents learn complex deceptive behaviours which is explained in more detail in the next subsection. We have described the different reward configurations used by us in more detail below:

\begin{itemize}[leftmargin=*]
    \item \textbf{Coverage Reward}: This reward is used by the good agents to learn how to collaboratively cover all the landmarks in the environment. The agents are rewarded on the basis of the bi-partite graph distance between the graph of agents and the graph of landmarks. In other words, agents would be given a higher reward if they are successfully covering the landmarks with 1:1 matching and would be given a lower reward if they are unable to so.
    \item \textbf{Deception Reward}: This is the reward which is used to actually learn deceptive behaviours. The reward given to the good agents is different than the reward given to the adversary. The good agents are rewarded on the basis of how close the adversary is to the target landmark. They are given a higher reward if the adversary is far away from target and a lower reward if it comes near the target landmark.
    
    The adversary's reward configuration is exact opposite to that of the good agents. We give a higher reward to the adversary if it gets nearer to the target landmark and give it a lower reward if it's unable to do so. Please note that this reward is only given to the adversary if it's learning. If we use a heuristic adversary, this reward configuration is used only for good agents and the adversary is not given any reward.
\end{itemize}

\begin{table}[t]
\renewcommand{\arraystretch}{1.5}
\caption{WEIGHTS USED TO TRAIN THE GOOD AGENTS}
\begin{center}
\label{reward_weights}
\begin{tabular}{ c | c | c }
\hline
INDEX & COVERAGE WEIGHT & DECEPTION WEIGHT \\
\hline
1 & 0.9 & 0.1 \\
\hline
2 & 0.8 & 0.2 \\
\hline
3 & 0.7 & 0.3 \\
\hline
4 & 0.6 & 0.4 \\
\hline
\end{tabular}
\end{center}
\end{table}

\begin{table*}[t]
\renewcommand{\arraystretch}{1.8}
\caption{SENSITIVITY ANALYSIS FOR 2 GOOD AGENTS VS ADVERSARY AGENT}
\begin{center}
\label{2_heuristic}
\begin{tabular}{c|c|c|c|c|c}
\hline
& \multicolumn{2}{c|}{\textbf{GOOD AGENTS}} & \multicolumn{3}{c}{\textbf{ADVERSARY AGENT}} \\
\hline
REWARD WEIGHT & BIPARTITE DIST. & DIST. THRESHOLD & TARGET DIST. & DIST. THRESHOLD & TARGET SELECT\\
\hline
0.9 cov, 0.1 dec & 0.12 $\pm$ 0.05 & 38.5 $\pm$ 5.3 & 0.77 $\pm$ 0.31 & 2.3 $\pm$ 3.7 & 19.8 $\pm$ 7.0 \\
\hline
0.8 cov, 0.2 dec & 0.13 $\pm$ 0.05 & 36.0 $\pm$ 6.0 & 0.76 $\pm$ 0.32 & 2.4 $\pm$ 4.8 & 18.0 $\pm$ 7.4\\
\hline
0.7 cov, 0.3 dec & 0.14 $\pm$ 0.05 & 29.2 $\pm$ 8.6 & 0.82 $\pm$ 0.31 & 1.5 $\pm$ 2.6 & 14.5 $\pm$ 5.5\\
\hline
0.6 cov, 0.4 dec & 0.70 $\pm$ 1.61 & 14.4 $\pm$ 9.5 & 0.85 $\pm$ 0.31 & 1.6 $\pm$ 3.4 & 11.6 $\pm$ 8.3 \\
\hline
\end{tabular}
\end{center}
\end{table*}

\begin{table*}[t]
\renewcommand{\arraystretch}{1.8}
\caption{SENSITIVITY ANALYSIS FOR 3 GOOD AGENTS VS ADVERSARY AGENT}
\begin{center}
\label{3_heuristic}
\begin{tabular}{c|c|c|c|c|c}
\hline
& \multicolumn{2}{c|}{\textbf{GOOD AGENTS}} & \multicolumn{3}{c}{\textbf{ADVERSARY AGENT}} \\
\hline
REWARD WEIGHT & BIPARTITE DIST. & DIST. THRESHOLD & TARGET DIST. & DIST. THRESHOLD & TARGET SELECT\\
\hline
0.9 cov, 0.1 dec & 0.11 $\pm$ 0.04 & 37.4 $\pm$ 8.1 & 0.93 $\pm$ 0.44 & 1.2 $\pm$ 1.9 & 11.3 $\pm$ 5.8\\
\hline
0.8 cov, 0.2 dec & 0.13 $\pm$ 0.05 & 30.0 $\pm$ 11.2 & 0.97 $\pm$ 0.46 & 1.3 $\pm$ 3.1 & 10.03 $\pm$ 6.4\\
\hline
0.7 cov, 0.3 dec & 0.13 $\pm$ 0.06 & 28.8 $\pm$ 8.0 & 0.94 $\pm$ 0.40 & 1.1 $\pm$ 2.0 & 9.3 $\pm$ 6.3 \\
\hline
0.6 cov, 0.4 dec & 0.37 $\pm$ 0.64 & 9.5 $\pm$ 9.2 & 0.94 $\pm$ 0.39 & 1.23 $\pm$ 2.4 & 8.5 $\pm$ 7.4 \\
\hline
\end{tabular}
\end{center}
\end{table*}

\subsection{Curriculum Learning}
We did some preliminary experiments using the deceptive reward described in the previous section. However, we noticed that the agents did not converge to any reasonable behaviour. So we propose a two step training process. We first train the good agents using the coverage reward. After they learn to cover all the landmarks with a high success rate, we gradually introduce a weighted deception reward. We observe in our experiments that the good agent assigned to the target landmark learns to stay a little away from the landmark instead of covering it in order to deceive the adversary. We performed a sensitivity analysis by varying the weights of deception and coverage and the results have been provided in the next section.

After training the good agents with the coverage reward, we train the agents with a weighted sum of deception and coverage reward. We did experiments with weights presented in Table \ref{reward_weights}.

\section{RESULTS AND DISCUSSION}
We performed experiments using the methods described and evaluated their performance in cases of 2 good agents and a heuristic adversary and 3 good agents and a heuristic adversary. This has been done by doing a sensitivity analysis over different weights of deception and coverage rewards.
\subsection{Evaluation Metrics}
We evaluated the performance using five different metrics. These are goal selected, bipartite distance, distance threshold and distance from target landmark. These metrics and evaluation methods for good agents and the adversary agent have been explained in more detail below.
\begin{itemize}[leftmargin=*]
    \item \textbf{Good Agents}: The following metrics have been observed and produced due to the behaviour displayed by the good agents:
    \begin{enumerate}
        \item \textbf{Bipartite Distance}: We calculate the mean bipartite distance between the good agents and the landmarks during an episode. We then calculate mean of these distances over 30 episodes. A lower mean bipartite distance would mean that the good agents are covering the targets properly.
        \item \textbf{Distance Threshold}: We calculate the number of steps during an episode for which bipartite distance of good agents is within a threshold of 0.1. We then calculate mean of these steps over 30 episodes.
    \end{enumerate}
    
    \item \textbf{Adversary Agent}: The following metrics have been observed and produced due to the behaviour displayed by the adversary agent:
    \begin{enumerate}
        \item \textbf{Distance from Target}: We calculate the mean of the distance between adversary and the target landmark during an episode. We then calculate mean of these distances over 30 episodes.
        \item \textbf{Distance Threshold}: We calculate the number of steps during an episode for which distance of adversary from target is within a threshold distance of 0.1. We then calculate mean of these steps over 30 episodes.
    \end{enumerate}
\end{itemize}

\subsection{Results}
We performed sensitivity analysis on 2 goods agents vs heuristic adversary and 3 good agents vs heuristic adversary. The corresponding results have been shown in Table \ref{2_heuristic} and Table \ref{3_heuristic}. We can notice that as we increase the deception reward weight in both cases, the number of times the target is selected by the adversary decreases. Also we can observe that the number of time steps for which the adversary agent comes near to the target landmark decreases with the increase in deception reward weight. All of this suggests that as we increase the weight of deception reward, the success rate of the good agents deceiving the adversary agent also increase. We also tried increasing the deception reward weight to 0.5 but in that case the good agents failed to converge to a reasonable behaviour.

One more interesting observation from the results can be that the bipartite distance between the good agents and the landmarks also increases as we increase the deception weights. This makes sense because as the deception weight increases, the good agent assigned to the target landmark in every episode learns to stay at some distance away from the target landmark. This result can be noticed visually also.

\subsection{Future Work}
In this work, we proposed augmenting the multi-agent reinforcement learning framework proposed by \cite{agarwal2019learning} with an opponent encoder module for learning multi-agent policies in presence of an adversary in the environment. We performed a two step training process with different reward configuration to train a team of agents to deceive an adversary in the environment.

Only preliminary results for different reward configurations against a heuristic agent have been presented in this work. A good future direction would be to investigate other reward configurations and to perform experiments with a learning adversary. The proposed framework can also be extended to learn multi-agent policies against a team of heuristic or learning adversaries. 




\section*{ACKNOWLEDGMENT}

 I would like to thank my wonderful set of collaborators Akshay Sharma, Dana Hughes, Fan Jia, Keitaro Nishimura, Sumit Kumar, Swaminathan Gurumurthy and Yunfei Shi for supporting me and for taking time out of their busy schedules for discussions about my work. This work has been funded in part by DARPA OFFSET award HR00111820029.

\bibliographystyle{IEEEtranS}
\bibliography{final_report}

\end{document}